\documentclass[twoside,journey]{IEEEtran}
\usepackage{makecell}
\usepackage{hyperref}
\usepackage{array}
\usepackage{graphicx,amssymb,amsmath}
\usepackage{multicol}
\usepackage[noadjust]{cite}
\usepackage{setspace}
\usepackage{subfigure}
\usepackage{graphicx}
\usepackage{float}
\usepackage{url}
\usepackage{stfloats}
\usepackage{amsthm,pifont}
\usepackage{cases,subeqnarray}
\usepackage{bm, multirow, bigstrut}
\usepackage{amsmath, amsthm, amssymb}
\usepackage{textcomp}
\usepackage{latexsym,bm}
\usepackage{booktabs}
\usepackage{xcolor}
\usepackage{colortbl}
\definecolor{lightskyblue}{RGB}{220, 230, 255} 
\definecolor{softorange}{RGB}{255, 230, 200}
\usepackage{mathtools}
\usepackage{dsfont}
\usepackage{extarrows}
\usepackage{multirow}
\usepackage{epsfig}
\usepackage{epsfig}
\usepackage{flushend}
\usepackage{epstopdf}
\usepackage[noend]{algpseudocode}
\usepackage{algorithmicx,algorithm}
\usepackage{newtxtext,newtxmath} 
\usepackage{helvet}              
\usepackage{courier}             
\usepackage{enumitem}
\theoremstyle{plain}

\theoremstyle{plain}

\usepackage{amsmath}

\IEEEoverridecommandlockouts
\begin{document}

\title{Towards Edge General Intelligence: Knowledge Distillation for Mobile Agentic AI}
\author{
Yuxuan Wu, Linghan Ma, Ruichen Zhang, Yinqiu Liu, Dusit Niyato,  \textit{Fellow, IEEE},\\Shunpu Tang, Zehui Xiong, \textit{Senior Member, IEEE}, Zhu Han, \textit{Fellow, IEEE}, Zhaohui Yang, \\Kaibin Huang, \textit{Fellow, IEEE}, Zhaoyang Zhang, \textit{Senior Member, IEEE}, Kai-Kit Wong, \textit{Fellow, IEEE}
\thanks{Y. Wu, L. Ma,  S. Tang, Z. Yang and Z. Zhang are with the College of Information Science and Electronic Engineering, Zhejiang University, Hangzhou, Zhejiang 310027, China(e-mail: \{3230100234, 3230100339, tangshunpu, yang\_zhaohui, zhzy \}@zju.edu.cn).}%
\thanks{R. Zhang, Y. Liu, and D. Niyato are with the College of Computing and Data Science, Nanyang Technological University, Singapore (e-mail: ruichen.zhang@ntu.edu.sg, yinqiu001@e.ntu.edu.sg, dniyato@ntu.edu.sg).}%
\thanks{Z. Xiong is with the School of Electronics, Electrical Engineering and Computer Science, Queen's University Belfast, Belfast, BT7 1NN, U.K. (e-mail: z.xiong@qub.ac.uk).}%
\thanks{Z. Han is with the University of Houston, Houston, TX 77004, USA, and also with the Department of Computer Science and Engineering, Kyung Hee University, Seoul 446701, South Korea (e-mail: hanzhu22@gmail.com).}%
\thanks{K. Huang is with the Department of Electrical and Electronic Engineering, The University of Hong Kong, Hong Kong SAR, China (e-mail: huangkb@eee.hku.hk).}%
\thanks{K. K. Wong are with the Department of Electronic and Electrical Engineering, University College London, WC1E 7JE, London, United Kingdom (e-mail: kai-kit.wong@ucl.ac.uk). K. K. Wong is also affiliated with Yonsei Frontier Laboratory, Yonsei University, Seoul, 03722, Republic of Korea.}
\thanks{Y. Wu and L. Ma contributed equally to the work.}
}
\maketitle
\begin{abstract}
Edge General Intelligence (EGI) represents a paradigm shift in mobile edge computing, where intelligent agents operate autonomously in dynamic, resource-constrained environments. However, the deployment of advanced agentic AI models on mobile and edge devices faces significant challenges due to limited computation, energy, and storage resources. To address these constraints, this survey investigates the integration of Knowledge Distillation (KD) into EGI, positioning KD as a key enabler for efficient, communication-aware, and scalable intelligence at the wireless edge. In particular, we emphasize KD techniques specifically designed for wireless communication and mobile networking, such as channel-aware self-distillation, cross-model Channel State Information (CSI) feedback distillation, and robust modulation/classification distillation. Furthermore, we review novel architectures natively suited for KD and edge deployment, such as Mamba, RWKV (Receptance, Weight, Key, Value) and Cross-Architecture distillation, which enhance generalization capabilities. Subsequently, we examine diverse applications in which KD-driven architectures enable EGI across vision, speech, and multimodal tasks. Finally, we highlight the key challenges and future directions for KD in EGI. This survey aims to provide a comprehensive reference for researchers exploring KD-driven frameworks for mobile agentic AI in the era of EGI.
\end{abstract}
\begin{IEEEkeywords}
    Edge General Intelligence (EGI), mobile agentic ai, Knowledge Distillation (KD), Wireless Edge Intelligence
\end{IEEEkeywords}

\IEEEpeerreviewmaketitle
\section{Introduction}\label{Intro}
\subsection{Background}
The Mobile Edge Computing market is undergoing substantial expansion, with projections indicating its value will surge from approximately USD 1.65 billion in 2024 to over USD 13.5 billion by 2032\footnote{https://www.maximizemarketresearch.com/market-report/global-mobile-edge-computing-market/6951/}. This growth is propelled by the increasing demand for low-latency computing and enhanced quality of experience (QoE) for a rapidly growing number of mobile and IoT devices. This evolving landscape is fostering a significant technological transformation known as ``agentification," where edge devices are being endowed with autonomous capabilities through the integration of Large language models (LLMs) and other advanced AI modules. This shift transforms passive edge nodes into proactive mobile agentic AI, systems capable of perceiving their environment, reasoning, and executing complex, multi-step tasks without direct human intervention.

The benefits of this agent-centric paradigm are manifold. mobile agentic AI \cite{ref4} enables a higher degree of automation and personalization, as agents can learn user preferences and adapt to dynamic situations in real-time, directly on the device. This on-device processing significantly reduces latency and enhances data privacy by minimizing reliance on centralized cloud servers. Furthermore, these agents can proactively anticipate issues, optimize workflows, and coordinate across different systems, leading to greater operational efficiency, faster issue resolution, and improved service agility. By embedding sophisticated cognitive abilities at the network's edge, mobile agentic AI paves the way for more intelligent, responsive, and secure applications, marking a critical step toward the realization of EGI, defined as the ability of edge devices to perform general-purpose reasoning and problem solving comparable to cloud AI under strict resource and latency constraints.\cite{ref2}.

The realization of EGI depends on the deployment of mobile agentic AI. LLMs provide the cognitive engine for such agents, demonstrating remarkable capabilities in planning, reasoning, and tool use. However, the immense computational, memory, and energy requirements of these LLMs are fundamentally incompatible with the resource-constrained nature of mobile and edge devices.\cite{int-y} This ``deployment chasm" is the primary obstacle to achieving the EGI vision.

To bridge this gap, KD has emerged as a key approach for compressing large models. KD refers to the process of training a smaller “student” model to mimic the behavior of a larger, more capable “teacher” model. This allows the student to preserve the teacher’s advanced capabilities in a compact form suitable for deployment on resource-constrained edge hardware \cite{ref5}.
\begin{table*}[th]
\centering
\caption{Summary of Related Surveys}
\label{tab:related_surveys}
\footnotesize
\renewcommand{\arraystretch}{1.4}
\begin{tabular}{|m{2cm}|m{0.3cm}|m{2.2cm}|m{10cm}|m{0.3cm}|m{0.3cm}|m{0.7cm}|}
\hline
\rowcolor{lightskyblue}
\textbf{scope} & \textbf{Ref.} & \textbf{Kernel} & \textbf{overview}&\textbf{EGI}&\textbf{KD}&\textbf{Agentic AI} \\
\hline
\multirow{4}{*}{EGI} & \cite{XU} &Edge Intelligence &A comprehensive survey with a 4 pillar framework about edge intelligence&$\checkmark$&$\times$&$\times$ \\ \cline{2-7}& \cite{ref2}& EGI via LLMs &A survey categorizing LLM-empowered EGI into centralized, hybrid, and decentralized systems and reviewing their implementations &$\checkmark$ &$\times$&$\times$ \\\cline{2-7} &\cite{Founda}&Foundation model towards EGI & A survey introducing foundation models as the key toward General Edge Intelligence,outlining research directions to tackle challenges &$\checkmark$ &$\times$&$\times$ \\\cline{2-7} & \cite{Zhao}&EGI with World Models &A survey introducing how world models can empower agentic AI with proactive planning and reasoning capabilities at the edge &$\checkmark$ &$\times$ &$\checkmark$\\
\hline
\multirow{2}{*}{Agentic AI}&\cite{ref1}&Agentic AI&A survey reviewing the transformative role of agentic AI in organizations, highlighting its core attributes
and the strategic shift&$\times$&$\times$&$\checkmark$\\ \cline{2-7}&\cite{ref4}&Mobile Agentic AI&A survey about multimodal mobile agents,
categorizing and comparing their deployment effectiveness on mobile devices&$\times$&$\times$&$\checkmark$\\
\hline
Other techniques&\cite{KDS}&Knowledge \newline distillation&A foundational survey on knowledge distillation, detailing its core components&$\times$&$\checkmark$&$\times$\\ \cline{2-7}
\hline
\rowcolor{softorange}
\multicolumn{2}{|c|}{Ours}&EGI with KD for mobile agentic AI& A comprehensive survey introducing KD as the key for mobile agentic AI towards EGI &$\checkmark$&$\checkmark$&$\checkmark$\\
\hline
\end{tabular}
\end{table*}
\subsection{Comparisons with Related Surveys and Contributions}
The framework of KD for advancing agentic AI and enabling its deployment in EGI has demonstrated substantial potential, thereby reducing computational overhead, and enhancing adaptability in dynamic wireless environments. This paper aims to provide a comprehensive survey on the fundamentals of KD, along with its applications in agentic AI, and to highlight the enabling technologies that open new avenues for EGI deployment. Table ~\ref{tab:related_surveys} presents an in-depth comparison of related surveys, emphasizing KD, agentic AI, and their implementation in EGI. These surveys have primarily focused on summarizing the evolution and optimization of KD. For example, Gou \textit{et al.} \cite{KDS} offered a foundational survey on KD, detailing its core components such as knowledge types, training schemes, and teacher-student architectures. 

Additionally, several surveys have investigated the application of agentic AI. Hosseini \textit{et al.} \cite{ref4} review the transformative role of agentic AI in organizations, highlighting its core attributes and the strategic shift from human-assisted ``Copilot" to ``autonomous Autopilot" models. Moreover, Wu \textit{et al.} \cite{ref1} survey the landscape of multimodal mobile agents, categorizing them into prompt-based and training-based methods and comparing their deployment effectiveness on mobile devices.

Furthermore, several studies have demonstrated the growing significance of agentic ai in shaping the evolution of EGI. Xu \textit{et al.} \cite{XU} provided a comprehensive survey on edge intelligence. As the development of powerful LLMs,the exploration of EGI have surged in recent times. Chen \textit{et al.} \cite{ref2} provide a foundational overview by categorizing LLM-empowered EGI into centralized, hybrid, and decentralized systems and reviewing their implementations. He \textit{et al.} \cite{Founda} proposed that the integration of foundation models is the key to evolving toward EGI, outlining research directions to tackle challenges. Focusing on the cognitive core of such systems, Zhao \textit{et al.} \cite{Zhao} offer a comprehensive analysis of how world models can empower agentic AI with proactive planning and reasoning capabilities at the edge.

However, existing studies lack sufficient investigation into systematic KD methodologies tailored to the constraints of EGI, as well as comprehensive analysis of resource limitations and model adaptation challenges faced by agentic AI in edge environments. This paper addresses these gaps by providing an in-depth examination of how advanced model adaptation techniques, such as wireless distillation \cite{Ac5,CSI2,CSI-ALM-Light,Gesture}, and some architectures designed for the edge \cite{ref55,ref56,ref60,ref61}. Furthermore, we demonstrate how KD can be effectively adapted to diverse EGI deployment scenariosn, such as autonomous vehicles \cite{ref134,ref135}, unmanned aerial vehicle (UAV) \cite{UAV1,UAV2}, robotics \cite{ref139,ref140}, and other Internet of Things \cite{IoT1,IoT2,IoT3}. The main contributions are summarized as follows.
\begin{itemize}
\item We provide a dedicated review of KD techniques in wireless communication scenarios, highlighting how KD enhances channel estimation, feedback compression, and resource-efficient model deployment at the wireless edge.
\item We systematically review existing KD techniques and their advantages, highlighting their potential integration with novel architectures. Afterward, we comprehensively discuss the benefits and opportunities of applying KD alongside these architectures in EGI.
\item We analyze the limitations present in existing models. Based on these shortcomings, we introduce architectures beyond Transformer and investigate tuning techniques combined with knowledge distillation to create conditions suitable for deployment on edge devices.
\item We further analyze the challenges currently faced by EGI from both technical and ethical dimensions, proposing potential future development trends and solutions.
\end{itemize}

\begin{figure}
    \centering
    \includegraphics[width=0.5\textwidth]{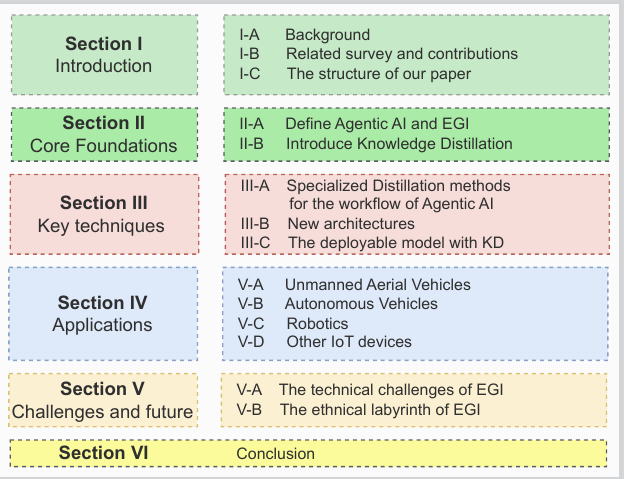}
    \caption{The structure of this survey}
    \label{structure}
\end{figure}
\subsection{The Structure of Paper}
The structure of this survey is outlined in Fig.~\ref{structure} Section~\ref{Core} introduces the fundamental concepts of EGI, mobile agentic AI, and KD, and further elucidates the interrelationships among these three technologies. Subsequently, Section~\ref{Three} systematically elaborates on novel architectures and wireless distillation that integrate KD into EGI, and presents the current advancements achieved by existing models. Section ~\ref{Four} delineate the role of KD in EGI and its application within specialized domains. Finally, Section~\ref{Five} elucidates the lessons learned from the review and challenges faced in this domain from both technical and macro-level perspectives, while outlining potential future research directions.

\section{Core Methodological Foundations}\label{Core}

Agentic AI provides an autonomous perception–planning–action–memory loop, while EGI extends this paradigm to the network edge by enabling generalized cognitive capabilities under strict resource constraints. Together, they outline a unified vision for intelligent, adaptive, and autonomous edge systems.

\subsection{The Rise of Agentic AI and the Vision of Edge General Intelligence}
\subsubsection{Mobile Agentic AI}

Agentic AI refers to autonomous systems that perceive, reason, plan, and act to achieve goals with minimal human supervision \cite{ref1}. These systems operate in a continuous cycle, often termed the agentic loop, which is comprised of four core components: Perception, Planning, Action, and Memory \cite{ref1}.
\begin{figure*}
    \centering
    \scriptsize
    \includegraphics[width=1.0\textwidth]{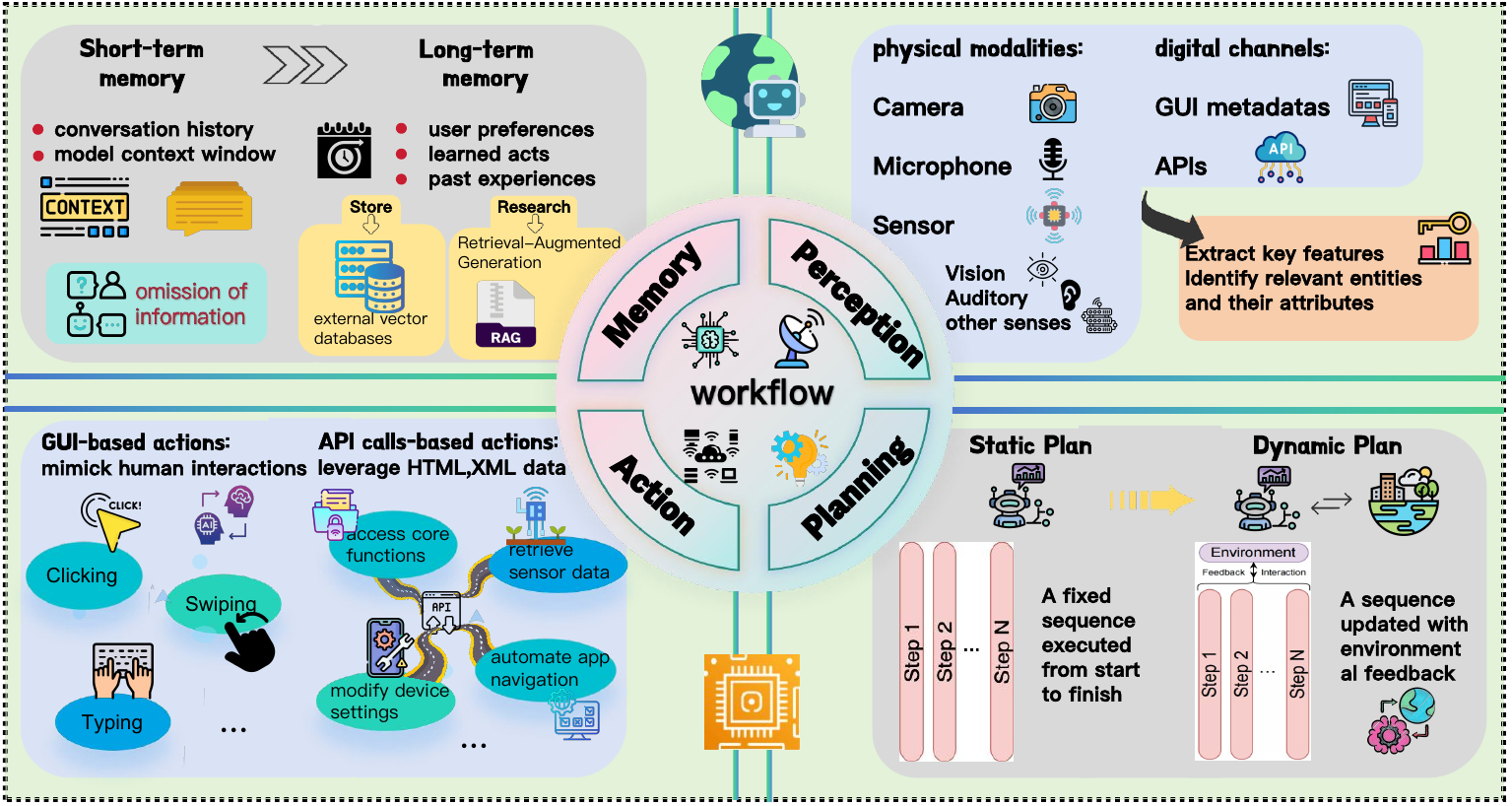}
    \caption{An overview of the workflow of Agentic AI. \textit{Perception} gathers and interprets multi-model information. \textit{Planning} devises a sequence of actions to achieve a high-level goal. \textit{Action} interacts with and affects its environment. \textit{Memory} enables an agent to retain information over time.}
    \label{EGI}
\end{figure*}
\begin{itemize}
\item \textbf{Perception}: The perception module integrates multimodal information to form a coherent understanding of the environment. As depicted in Fig.~\ref{EGI}, this process begins with data acquisition from physical sensory streams or digital channels. Subsequently, these raw data are processed to extract key features and identify relevant entities and their attributes.

\item \textbf{Planning}: The planning module devises a sequence of actions to fulfill a high-level goal, frequently leveraging LLMs as a cognitive core. As Fig.~\ref{EGI} shows, agentic planning shifts from static, predefined action sequences to dynamic plans that are continuously updated based on environmental feedback, ensuring robust adaptability.

\item \textbf{Action}: The agent executes the chosen action, interacting with and affecting its environment. Actions can range from mimicking human interactions with graphical user interfaces (GUIs), such as clicking or typing \cite{ME} in Fig.~\ref{EGI}, to leveraging API calls for deeper system integration, such as modifying device settings or automating app navigation. In multi-agent systems, communication itself becomes a pivotal action, enabling agents to coordinate and delegate tasks \cite{ME}.

\item \textbf{Memory}:The memory module allows agents to retain information over time, providing essential context for coherent interactions and facilitating continual learning. As shown in Fig.~\ref{EGI}  This is typically achieved in two tiers:

Short-term memory preserves the context of the current session, often managed within the LLM's context window.
Long-term memory stores knowledge across sessions, such as learned facts and past experiences, commonly implemented using external vector databases. Agents can query this knowledge base via similarity search, a core mechanism of Retrieval-Augmented Generation (RAG).

\end{itemize}

Moreover, Coordination and adaptation are critical capabilities that enable Agentic AI to manage complex, long-horizon tasks in dynamic environments \cite{ADAP,COOD}. Coordination is achieved through dynamic multi-agent collaboration, where autonomous agents interact to reach shared objectives \cite{COOD}. Complementing this, adaptation is the capacity of an agent to learn from its environment and modify its behavior in real-time to pursue high-level goals \cite{ADAP}. This ability is fundamentally underpinned by integrated memory systems and learning mechanisms such as reinforcement learning \cite{RL} and lifelong learning \cite{LL2}. The synergy between robust coordination protocols and memory-driven adaptation distinguishes modern Agentic AI, paving the way for more resilient and autonomous systems.

When deployed in wireless and edge computing environments, each component of the agentic loop must operate under constraints of limited computation, storage, and communication bandwidth. This challenge has catalyzed a shift in communication protocol design, moving from a focus on raw throughput to ``semantic efficiency" and prioritizing the transmission of concise, high-value information over verbose raw data \cite{WL}. Emerging frameworks for 6G envision ``content-aware" networks that can intelligently prioritize semantically critical data, blurring the lines between the application and network layers and transforming the network from a passive conduit into an active participant in information exchange \cite{WL}.

\subsubsection{Edge General Intelligence}

EGI \cite{ref2} represents a transformative paradigm designed to endow edge devices with general-purpose cognitive capabilities, enabling them to perceive, reason, and act autonomously in dynamic environments. The ultimate vision of EGI is to achieve artificial general intelligence (AGI) at the network edge, where systems can attain high-level cognitive abilities such as comprehension, reasoning, planning, and learning from experience at or even beyond human-level proficiency.

Unlike traditional edge intelligence, which primarily deploys static, task-specific models for individual tasks, EGI emphasizes versatility, adaptability, and autonomous cognitive reasoning \cite{XU}. To achieve this, EGI leverages foundation models, enabling devices to perform multiple diverse tasks without frequent retraining and to dynamically adapt to varying contexts and environments in real-time. 

This intelligence is fundamentally rooted in the knowledge, a concept that encompasses not just factual information but also complex reasoning patterns, contextual understanding, and decision-making abilities learned from vast datasets, embedded within AI models.
EGI architectures exist on a spectrum defined by how this knowledge is distributed:
\begin{itemize}
\item In a centralized model, a single, powerful, cloud-hosted LLM possesses all the comprehensive knowledge. It handles all complex reasoning and planning, while edge devices, possessing minimal local knowledge, serve as simple data collectors and command executors.

\item Conversely, a fully decentralized approach aims to imbue each edge device with its own substantial knowledge base by equipping it with a capable medium language model or small language model (SLM), enabling on-device inference and peer-to-peer collaboration.\cite{dec}

\item A hybrid architecture balances these extremes, using local SLMs with specialized or essential knowledge for routine and latency-sensitive operations, while offloading tasks requiring broader or more profound knowledge to the cloud LLM.
\end{itemize}
The potential applications for EGI are vast and transformative, promising to redefine human-machine interaction across numerous sectors, including autonomous vehicles, industrial automation, smart cities, and personalized healthcare.

\begin{figure*}
    \centering
    \includegraphics[width=0.8\textwidth]{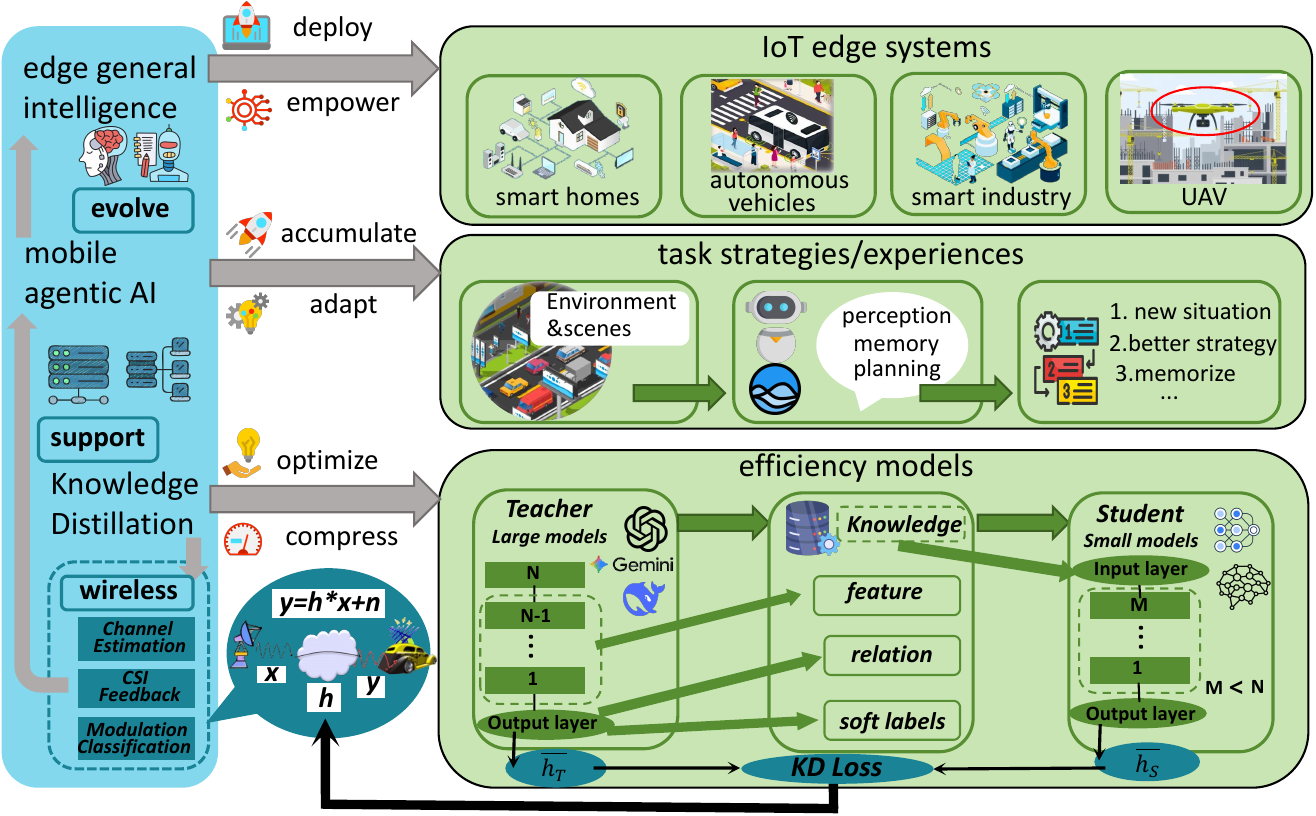}
    \caption{Knowledge Distillation compresses large models into lightweight ones for deployment, enabling Mobile Agentic AI to accumulate experiences and adapt strategies. These capabilities collectively foster Edge General Intelligence, which empowers IoT edge systems such as UAVs, autonomous vehicles, and smart healthcare.}
    \label{bridge}
\end{figure*}

\subsection{Knowledge Distillation: Core Techniques and Pardigms}

KD provides an effective framework for transferring knowledge from large models to compact ones, enabling efficient deployment on resource-constrained devices. Modern KD methods based techniques—capture increasingly richer forms of teacher knowledge, forming the foundation of high-performance edge and agentic AI systems.

\subsubsection{Basic Concepts and Working Principles}

Hinton \textit{et al.} \cite{ref5} firstly proposed the teacher-student architecture, constituting the foundational paradigm of KD. This architecture typically integrates two components: (i) a high-capacity teacher model, frequently implemented as a state-of-the-art neural network or model ensemble, that generates rich knowledge representations and (ii) a compact student model optimized for efficient deployment. Through a specialized distillation process, the student model learns to replicate the teacher's functional mapping by approximating its knowledge outputs.

The core of this framework is the knowledge transfer mechanism \cite{ref5}. The teacher model provides not just a single correct answer but a full probability distribution across all possible classes. These nuanced probabilities, often called ``soft targets" or ``dark knowledge," \cite{KD1} encode rich information about how the teacher generalizes and perceives similarities between classes, providing far richer supervision than the ``hard," one-hot labels used in standard training. To produce these effective soft targets, a temperature scaling parameter \(T\) \cite{KD2} is applied to the teacher's output layer, so that the probability assigned to class \(i\) is given by
\begin{equation}
q_i = \frac{\exp(z_i / T)}{\sum_j \exp(z_j / T)} ,
\end{equation}
where  $q_i$ denotes the probability assigned to class  $i$, $ z_i$ represents its corresponding logit output,
and  $T$  represents the temperature.

The overall training objective of the student model consists of two components. (i) loss function with soft targets: The student model employs the same high temperature \( T \) as the teacher in its softmax layer to compute the cross-entropy loss between its output distribution and the soft targets provided by the teacher. (ii) loss function with hard targets: In parallel, the student is also trained to predict the ground-truth labels from the original training data, i.e., the hard targets. For this component, the softmax temperature is set to 1 to reflect standard classification behavior. The combined loss can be written as
\begin{equation}
\mathcal{L} = \alpha \cdot \mathcal{L}_{\text{soft}} + (1 - \alpha) \cdot \mathcal{L}_{\text{hard}},
\end{equation}
where
\begin{equation}
\mathcal{L}_{\text{soft}} = \text{L}\left(\sigma\left(\frac{\mathbf{z}^S}{T}\right), \sigma\left(\frac{\mathbf{z}^T}{T}\right)\right),
\end{equation}
and
\begin{equation}
\mathcal{L}_{\text{hard}} = \text{L}\left(\sigma(\mathbf{z}^S), \mathbf{y}\right).
\end{equation}
Here, $\mathbf{z}^S$ and $\mathbf{z}^T$ denote the logits of the student and teacher models, respectively, and $\mathbf{y}$ is the one-hot encoded ground-truth label. Let $\mathbf{p}^S = \sigma(\mathbf{z}^S / T)$ and $\mathbf{p}^T = \sigma(\mathbf{z}^T / T)$ represent the softened probability distributions, where $\sigma(\cdot)$ denotes the Softmax function and $T$ is the temperature parameter. 

The distillation objective $L$ is defined as the Kullback--Leibler (KL) divergence, which measures the discrepancy between the teacher's and student's distributions:
\begin{equation}
L_{\mathrm{KL}}(\mathbf{p}^T \,\|\, \mathbf{p}^S) 
= \sum_{i=1}^{C} p_i^T \log \left( \frac{p_i^T}{p_i^S} \right).
\end{equation}

Alternatively, when adopting the Mean Squared Error (MSE), the loss is formulated as:
\begin{equation}
L_{\mathrm{MSE}}(\mathbf{p}^T, \mathbf{p}^S)
= \frac{1}{C} \left\| \mathbf{p}^T - \mathbf{p}^S \right\|_2^2.
\end{equation}

Finally, $\alpha \in [0,1]$ is a hyperparameter that balances the weight between the ground-truth cross-entropy loss and the distillation loss term $L$.

\begin{table*}[h!]
\tiny
    \caption{A Comparative Analysis of Three Knowledge Distillation Paradigms}
    \centering
    \label{KD contrast}
    \resizebox{\textwidth}{!}{%
    \begin{tabular}{m{2cm}||m{2cm}|c|m{2cm}|m{2cm}|m{4cm}|m{4cm}}
    \hline
    \hline
    \rowcolor{lightskyblue}
    \textbf{Category} & \textbf{Representative Method} & \textbf{ref} & \textbf{Distillation Target} & \textbf{Layer Level} & \begin{center}\textbf{Pros}\end{center} & \begin{center}\textbf{Cons} \end{center}
    \\
    \hline
    Response-based & teacher-student & \cite{ref4} & softened logits (class probability) & output layer &
\parbox[c]{3.5cm}{ 
   \vspace*{5pt} 
    \begin{itemize}[leftmargin=*, noitemsep, topsep=0pt, partopsep=0pt]
        \item Conceptually simple
        \item Easy to implement
    \end{itemize}
    \vspace*{5pt} 
}
&
\parbox[c]{3.5cm}{
\vspace*{5pt} 
\begin{itemize}[leftmargin=*, noitemsep, topsep=0pt, partopsep=0pt]
        \item Restricted to output-layer information
        \item Neglect intermediate feature representations
        \item Limited capacity to transfer structural or semantic knowledge
    \end{itemize}
    \vspace*{5pt} 
} \\
\hline
Feature-based & FitNets & \cite{ref34} & intermediate feature maps & hidden layer &
\parbox[c]{3.5cm}{
\vspace*{5pt} 
     \begin{itemize}[leftmargin=*, noitemsep, topsep=0pt, partopsep=0pt]
        \item Leverages rich intermediate representations
        \item Enables the transfer of structural and semantic information
    \end{itemize}
    \vspace*{5pt} 
} &
\parbox[c]{3.5cm}{
\vspace*{5pt} 
    \begin{itemize}[leftmargin=*, noitemsep, topsep=0pt, partopsep=0pt]
        \item Challenging alignment between heterogeneous architectures
        \item High computational and memory costs 
    \end{itemize}
    \vspace*{5pt} 
} \\
\hline
Relation-based & Relational Knowledge Distillation(RKD) & \cite{ref41} & pairwise/triplet relational distance & across multiple layers &
\parbox[c]{3.5cm}{
\vspace*{5pt} 
    \begin{itemize}[leftmargin=*, noitemsep, topsep=0pt, partopsep=0pt]
        \item Captures higher-order dependencies among samples
        \item Enhances generalization
    \end{itemize}
    \vspace*{5pt} 
} &
\parbox[c]{3.5cm}{
\vspace*{5pt} 
    \begin{itemize}[leftmargin=*, noitemsep, topsep=0pt, partopsep=0pt]
        \item Computationally intensive and complex
        \item Sensitive to data distribution or noisy relations
    \end{itemize}
    \vspace*{5pt} 
} \\
\hline
\hline
    \end{tabular}
    }
\end{table*}
\subsubsection{A taxonomy of modern KD techniques}
  
KD methods based on knowledge sources can be broadly classified into three types: response-based distillation, feature-based distillation and relation-based distillation.

\paragraph {\bf{Response-Based Distillation}}

Response-based KD transfers knowledge by using the teacher’s final outputs to supervise the student model. The earliest teacher–student formulation of KD was inherently response-based \cite{ref5}.

Building on the KD framework, several variants have been developed to address its different limitations, forming a progressive line of research. Together, these works illustrate a spectrum of strategies from bridging model capacity gaps, to replacing or removing teachers, to designing more robust and adaptive loss functions. Mirzadeh \textit{et al.} \cite{ref6} extended the teacher–student paradigm with an assistant model to mitigate the capacity gap between a large teacher and a small student. Moreover, their framework outperformed other strong baselines—including FitNets, Attention Transfer (AT) and Mutual Learning. This yielded more effective transfer under constrained resources, though at the cost of extra training stages. Building on the idea of removing external teachers, Pham \textit{et al.} \cite{ref8} proposed self-distillation, where a model distills knowledge from its own intermediate predictions. This design is attractive for continual on-device learning, though its effectiveness hinges on prediction quality. Empirical results show that self-distillation consistently improves over teacher baselines, with student models achieving up to 1–2.5 percentage points higher accuracy on CIFAR-10/100 under comparable settings. Taken together, these works trace a clear trajectory, from bridging teacher–student gaps, to eliminating the need for teachers altogether, to strengthening robustness under noisy conditions. The collective insight suggests future research will likely integrate these dimensions, seeking hybrid methods that are resource-efficient, adaptive, and robust, while addressing scalability and parameter sensitivity. When viewed through the lens of agentic AI and mobile deployment, these directions highlight KD’s potential at the edge.

\paragraph{ \bf{Feature-Based Distillation}}

Feature-based knowledge distillation leverages intermediate feature representations of the teacher network to guide student training \cite{ref36}. The basic paradigm typically consists of two stages \cite{ref34,ref35,ref36,ref37}:

The first stage is Intermediate Feature Alignment. In this stage, a teacher layer is selected as the hint layer, and a corresponding student layer as the guided layer. A regressor is introduced to project the features from the student’s guided layer into the representation space of the teacher’s hint layer. The loss for this stage is defined as
\begin{equation}
L_{HT} = L\left(u_h(x) - r\left(v_g(x)\right)\right),
\end{equation}
where \( L_{HT} \) is the loss function, \( L \) is a custom metric, \( u_h(x) \) is the output of the teacher's hint layer, \( v_g(x) \) is the output of the student's guided layer, and \( r \) is the regressor function.

The second stage is KD Phase. In this stage, the aligned student is further trained with ground-truth labels and the teacher’s soft labels, while parameters learned in the first stage remain fixed.

Several extensions have further enriched the KD paradigm by deepening teacher–student interactions. Gou et al. \cite{ref38} proposed Reciprocal Teacher–Student Learning, where knowledge transfer is bidirectional. The student not only learns from the teacher but also provides feedback for adaptive teaching. This design increases adaptability but requires careful coordination of the mutual updates. Chen \textit{et al.} \cite{ref39} advanced supervision quality through Knowledge Review, which leverages shallower or multiple teacher layers to supply more diverse guidance signals, thereby alleviating the risk of overfitting to a single teacher representation. Complementing these supervision strategies, Guo \textit{et al.} \cite{ref40} developed attention-based distillation, encouraging the student to mimic both feature activations and teacher attention maps, thus capturing structural dependencies beyond raw outputs. The progression of these works highlights a growing trend of moving beyond output-level mimicry toward exploiting deeper representational structures for more effective knowledge transfer.

\paragraph{ \bf{Relation-Based Distillation}}

Relation-based distillation transfers pairwise similarity information that captures the structural relationships among samples \cite{ref42}. Its core design relies on a relation potential function, typically defined using distance-based potentials or angle-based measures \cite{ref43,ref47}.

Several variants further refine the KD paradigm by focusing on structural information and feature relationships. Tian \textit{et al.} \cite{ref45} introduced a contrastive learning framework, where student and teacher embeddings are aligned via positive and negative sample pairs. This shifts distillation toward modeling structural relationships rather than solely mimicking outputs. Building on the idea of preserving structure, Liu \textit{et al.} \cite{ref46} developed Inter-Channel Correlation Distillation, which enforces channel-wise correlation consistency. This balances feature diversity and stability, but may introduce computational overhead in high-dimensional channels. Extending structural preservation to the relational domain, Xin \textit{et al.} \cite{ref48} proposed Neighborhood Relation KD (NRKD), which distills local neighborhood relationships instead of relying on global similarity, thereby improving robustness to noisy or imbalanced data. This progression reflects a broader shift in KD research—from simple output matching toward more nuanced modeling of representation structures at multiple levels of granularity.

\subsection{Knowledge Distillation as the Bridge to Mobile Agentic AI and Edge General Intelligence}

Agentic AI is central to EGI, enabling edge devices to act as proactive, goal-driven systems rather than passive executors. However, agentic capabilities are inherently procedural (``how to think"), not merely static knowledge (``what to say"), creating a mismatch with conventional model compression methods that focus on reproducing output probabilities. Such techniques imitate final predictions but fail to capture the underlying reasoning process required for agentic behavior.

KD emerges as a key technology for reconciling the computational demands of large-scale AI with the limited resources of edge devices. As illustrated in Fig.~\ref{bridge}, it acts as a conceptual bridge enabling both mobile agentic AI and EGI by making advanced intelligence feasible on edge platforms. Through distilling massive, multi-billion-parameter “teacher’’ models into compact “student’’ models, KD substantially reduces model size, memory usage, and computation cost. This compression also improves operational efficiency: lightweight models deliver lower inference latency, essential for real-time EGI tasks such as autonomous navigation and industrial robotics, and significantly reduce energy consumption, extending the battery life of edge devices. Moreover, on-device inference enhances privacy and reliability by keeping sensitive data local and reducing dependence on unstable network connections.

In the context of Agentic AI, it is crucial to move beyond the traditional view of KD as mere compression and redefine it as a framework for capability transfer. The ultimate objective is not simply to reduce model size but to comprehensively transfer the complex cognitive skills essential for agentic behavior \cite{ref4}. Key challenges arise in this process, including retaining generalization under resource-constrained mobile environments\cite{Cha1}, preserving robustness against noisy or dynamic data, and reducing sensitivity to hyperparameters. Emerging solutions involve leveraging KD in wireless scenarios\cite{CSI2} to enable efficient over-the-air learning, designing lightweight and adaptive frameworks tailored for edge deployment, and developing tuning techniques that align well with KD. These lessons highlight KD’s evolving role from compression to capability transfer, paving the way toward resource-efficient and resilient agentic intelligence on mobile platforms.\cite{Cha2}
  
\section{Distilling Agentic Capabilities for Mobile Deployment}\label{Three}

KD establishes a layered relationship with EGI. At the capability level \cite{ref144}, KD equips EGI with fundamental capabilities, such as perception, planning, action and memory, that are required for robust operation and completion. At the methodological level \cite{Cap1}, a range of frameworks and techniques aligned with KD enhance computational efficiency and reduce resource consumption \cite{Cap4}, while simultaneously fostering generalizable representations that better align with diverse downstream tasks in EGI. At the toolkit level, KD itself offers reliable compact models that directly support the deployment of EGI. \cite{Cap3} This section proceeds to elaborate on how KD can be leveraged for EGI deployment across the three levels.

\subsection{Specialized Distillation Frameworks for Capabilities of Agentic AI}
To address these limitations, recent research has focused on developing specialized distillation frameworks tailored for agentic systems. A key principle unifying these advanced methods is a shift from imitating raw outputs toward transferring more abstract, structured, and reusable knowledge components.

\begin{table*}[h!]
\scriptsize
\centering
    \caption{KD for Wireless Applications}
    \label{KD_wireless}
    \begin{tabular}{m{2.5cm}||m{4cm}|m{6.5cm}|m{3.5cm}}
    \hline
    \hline
    \rowcolor{lightskyblue}
    \textbf{Application Task} & \textbf{Role of KD} & \begin{center}\textbf{Improvements}\end{center} & \begin{center}
        \textbf{Implications for EGI}
    \end{center} \\
    \hline
    Channel Estimation & 
   
      Teacher networks provide ``lossless gradients'' and
     Defensive distillation enhances robustness
 & 

$\bullet$  Higher estimation accuracy\newline
$\bullet$  Improved security (lower ASR from 0.9 to 0.06 \cite{ES2})\newline
$\bullet$ Overcomes non-differentiable training bottlenecks (A sum rate of 9.0 at 30 dB SNR, compared to 8.2 without KD \cite{ES3})
 & Provides reliable and robust channel modeling foundation \\
    \hline
    \rowcolor{softorange}
    CSI Feedback & 
    To  deploy complex CSI feedback models on resource-limited
    devices
 &
 \textbf{Improved model's performance} \newline
 $\bullet$ Up to 7.6 dB in NMSE \cite{CSI1} \newline
$\bullet$  NMSE dropping from -9.54 dB to -9.75 dB \cite{CSI2} \newline
$\bullet$ improvement of 0.40–0.54 dB PSNR \cite{CSI3}) \newline
 \textbf{Reduced training time and computational complexity} \newline $\bullet$ Reduced computational complexity to 25.50\%-43.46\% \cite{CSI1} \newline
$\bullet$ Reduced total training time by 66.8\% \cite{CSI2}
& Enables efficient inference at edge nodes \\
   \hline
    Modulation classification & Compresses complex detection networks while retaining generalization & 

    \vspace{1ex}
    \textbf{Higher accuracy} \newline
$\bullet$ Boosted the student’s maximum accuracy by 2.4\% \cite{MC1}\newline
$\bullet$ 77.86\% accuracy under PGD attack at 30 dB SNR \cite{MC2}\newline
 &Enables efficient inference at edge nodes \\
    \hline
    \rowcolor{softorange}
 Beam prediction & 
   Transfers knowledge from multi-sensor model to lightweight and enhances beam prediction model's robustness and reliability
 & 
$\bullet$ Achieved a lower Mean Squared Error under attack \cite{PL1}\newline
$\bullet$ Achieves 86.33\%of the teacher's accuracy with only 12.4\% of the parameters \cite{PL2}
 &
Enables the deployment of a secure and robust beam prediction model on base stations \\
    \hline
    
    Resource Allocation & 
 
   Transfers prior knowledge via teacher models to reduce search space
& 
$\bullet$ Reduced Energy Consumption by an average of 94\% compared to standard Federated Learning \cite{RDKD2} \newline
$\bullet$ Lower Latency and Cost \cite{RDKD}\newline
$\bullet$ Improved Accuracy with the algorithm converging faster to accuracy level of approximately 0.80 \cite{RDKD}
   &
    Facilitates efficient resource coordination for EGI \\
    \hline
    \hline
    \end{tabular}
\end{table*}
\subsubsection{Capability in``Perception"}
KD is crucial for establishing the perception capabilities of Agentic AI. An agent's awareness of the wireless environment depends on precise CSI estimation, accurate signal classification and so on. To overcome the difficulty of large model for direct deployment, KD enables efficient, accurate, and robust environmental sensing on resource-constrained edge devices.\cite{Per1,Sen}
\begin{itemize}
\item \textbf{Channel Estimation:} Knowledge-driven deep learning models, such as the network developed by Yang \textit{et al.} \cite{ES1} that integrates the knowledge of prior communication systems into its architecture, are increasingly being used for complex channel estimation. To address the practical deployment challenges, KD has become a key enabling technique. Catak \textit{et al.} \cite{ES2} employed a defensive distillation framework to improve a model's resilience to malicious signal perturbations, smoothing the decision surface and drastically reduces the success rate of attacks. This method is outstanding with its security enhancement by lowering the Attack Success Ratio (ASR) from approximately 0.9 down to 0.06. Kong \textit{et al.} \cite{ES3} used KD to solve a fundamental training problem where a non differentiable binarization layer for channel quantization provides inaccurate gradients and hinders learning. An auxiliary ``teacher" network was introduced to bypass this layer and provide ``lossless gradients", guiding the main ``student" network to a better solution. This technique gained a novel solution to a core optimization challenge, achieving a sum rate of approximately 9.0 at 30 dB Signal-to-Noise Ratio (SNR), compared to 8.2 without KD.
\item \textbf{CSI Feedback:} 
To deploy complex Channel State Information (CSI) feedback models on resource-limited devices, Tang \textit{et al.} \cite{CSI1} used KD to create lightweight models without difficult manual redesign. The method improved the student's performance by up to 7.6 dB in Normalized Mean-Squared Error (NMSE) while reducing the encoder's computational complexity to 25.50\%-43.46\% of the teacher's. However, a limitation is that the student's performance cannot surpass the teacher's. Cui \textit{et al.} \cite{CSI2} advanced this by proposing an "Encoder KD" framework to reduce the high training overhead of full autoencoder distillation, with the lightweight student encoder trained via KD before being combined with the powerful teacher decoder. This reduced total training time by 66.8\% and further improved feedback performance, with NMSE dropping from -9.54 dB to -9.75 dB. Beyond efficiency, Gong \textit{et al.} \cite{CSI3} used KD to enhance robustness against imperfect CSI in MIMO semantic communication to solve a problem where the ambiguous relationship between estimated and true CSI hinders model convergence. A teacher with perfect CSI guided a student trained on imperfect CSI, helping it learn a resilient representation. This KD stage alone contributed to an overall improvement of 0.40–0.54 dB Peak Signal-to-Noise Ratio (PSNR) over benchmarks including DeepJSCC \cite{CB1}, DeepJSCC-V \cite{CB2}, DeepJSCC-MIMO \cite{CB3}, and SwinJSCC \cite{CB4}, which are pre-trained with accurate CSI and then finetuned with estimated channel matrix.
\item \textbf{Modulation Classification:} 
To deploy complex Automatic Modulation Classification (AMC) models on edge devices, Yang \textit{et al.} \cite{MC1} utilized KD to improve the classification accuracy of the computationally simple student model without increasing its complexity. The method boosted the student's maximum accuracy by 2.4\% while the student's floating point operations remained less than half of the teacher's. Further enhancing reliability, Xu \textit{et al.} \cite{MC2} proposed an adversarial robust distillation strategy to transfer the resilience of a large, adversarially-trained teacher to a compact student. This was necessary because lightweight models are more vulnerable to attacks. Their proposed KD method enabled the student model to achieve 77.86\% accuracy under a Projected Gradient Descent (PGD) attack at 30 dB SNR, which was 4.6\% higher than AT and about 30\% higher than other robust distillation techniques.
\end{itemize}
This line of work is central to the Perception stage of an agentic AI workflow. Through KD, a compact student model can be deployed on edge devices to enable real-time and reliable perception of the wireless environment. Accurate channel-state understanding forms the foundation for the agent’s higher-level cognitive functions, supporting autonomous and secure operation in dynamic settings.

\subsubsection{Capability in ``Planning"}

By using knowledge distillation, the complex and computationally intensive planning processes of a large teacher model, which encompass procedural reasoning and high-level strategies, can be compressed into a compact and efficient student model suitable for real-world deployment.

\textit{Kuzlu et al.} \cite{PL1} applied a defensive distillation framework to transfer a teacher model's robustness to a compact student, improving the beam prediction model's resilience against adversarial attacks. \textit{Park et al.} \cite{PL2,PL3} introduced cross-modal relational KD to transfer knowledge from a complex multimodal teacher with the use of LiDAR, radar, GPS, and RGB camera to a lightweight radar-only student, enabling resource-efficient beam prediction without relying on expensive sensor data. Moreover, \textit{Yang et al.} \cite{PL4} developed a knowledge-driven approach by unrolling the domain knowledge of the iterative algorithm into a Graph Neural Network (GNN), creating a scalable and interpretable model for efficient resource allocation. Furthermore, \textit{Gong et al.} \cite{PL5} employed distillation methods to transfer a policy learned by a computationally expensive deep reinforcement learning (DRL) agent into a gradient boosting decision tree model, which achieves faster and more cost-effective resource optimization.

A robust and efficient planning faculty acts as the agent's cognitive engine, translating its goals and perceived state into a structured and executable course of action. This is the foundational layer that enables the agent to operate with foresight and intentionality, facilitating sophisticated, goal-directed behavior in dynamic and complex environments.

\paragraph{\bf{Policy Distillation}}
Policy distillation operates on the core principle of behavioral mimicry. The ``knowledge" transferred is the teacher's decision-making policy, typically represented as a probability distribution over all possible actions for a given state \cite{PKD}.

To dynamically manage resources in network slicing, Li \textit{et al.} \cite{PD1} used DRL to create policies that adapt to volatile user demand without a traffic model. This approach could learn effective policies for complex environments, but is limited by the high training cost.  To efficiently implement these policies in distributed settings, policy distillation has emerged as a critical technique to optimize decision-making models in wireless networks, enabling efficient and collaborative resource management. \textit{Zhou et al.} \cite{PD2} developed a federated bidirectional distillation mechanism to reduce the high communication overhead inherent in Federated Reinforcement Learning. By distilling knowledge to and from a central server, the method cut the required communication rounds by nearly 50\% in non-IID scenarios. Moving beyond simple compression, \textit{Mensah et al.} \cite{PD3} introduced a federated mutual policy distillation scheme for collaborative resource trading among heterogeneous agents. This was necessary to overcome learning instability and non-convergence caused by conflicting objectives and non-IID data in multi-agent systems. By allowing agents to learn from each other's policies, the method improved the total system utility by 12.5\% compared to baseline DRL techniques.

Despite its effectiveness, policy distillation exhibits inherent limitations. As a quintessential ``closed-box" method, it transfers final decisions while the underlying reasoning and causal logic are lost \cite{black}. Furthermore, if the environment changes, a mismatch between the distilled policy and the value function can lead to incorrect judgments, training divergence, and catastrophic forgetting \cite{black2}.

\paragraph{\bf{Interpretable Strategy Teaching}}
The central idea of Interpretable Strategy Teaching is to extract and transfer the knowledge embedded in a teacher agent in the form of explicit, interpretable, natural language strategies \cite{int1}.

To enhance the interpretability of AI in wireless communications, \textit{Zaidi et al.} \cite{INWL1} pioneered an intrinsic approach by developing a domain knowledge-guided neural network whose architecture directly reflects the physics of radio propagation, making its decisions inherently transparent and robust. In contrast, \textit{Chen et al.} \cite{INWL2} adopted a post-hoc strategy for complex DRL agents, proposing a framework that analyzes an agent's behavior to infer comprehensible decision heuristics, thereby building trust and providing actionable insights for engineers.

This paradigm offers several advantages. Natural language strategies improve interpretability and enable expert auditing \cite{int3}. It is also compatible with closed-box models, requiring only API access rather than internal parameters \cite{int3}. By separating a general-purpose model from an editable strategy library, it yields a modular hybrid intelligence framework that supports lifelong learning and enhances the safety and trustworthiness of agentic AI systems.

\paragraph{\bf{Chain-of-Thought (CoT) Distillation}}
CoT prompting \cite{CoT} represents a pivotal technique for generating the reasoning component of an agent's trajectory. CoT guides a large language model to emulate a human cognitive process by generating intermediate reasoning steps prior to reaching a final answer \cite{step}. The technique augments few-shot exemplars in a prompt with explicit reasoning chains, formatting examples as (input, CoT, output) rather than simple (input, output) pairs. \textit{Wang et al.} \cite{CoTWL} proposed a framework where a CoT enabled LLM translates high-level user intents into executable wireless network control policies. They used CoT to overcome the key limitations of standard LLMs in wireless applications, such as their poor multi-step reasoning, lack of interpretability, and tendency to generate incorrect or "hallucinated" outputs. This resulted in significant, measurable gains, including a 27.2\% increase in network sum rate in a Unmanned Aerial Vehicle (UAV) case study compared to a non-CoT baseline.

CoT-based KD is a methodology designed to transfer the complex, multi-step reasoning capabilities of large "teacher" language models to smaller, more efficient "student" models, enabling the model to decompose complex problems into manageable sub-problems for sequential resolution \cite{CoTKD}. 

\begin{figure}
\centering
\includegraphics[width=0.5\textwidth]{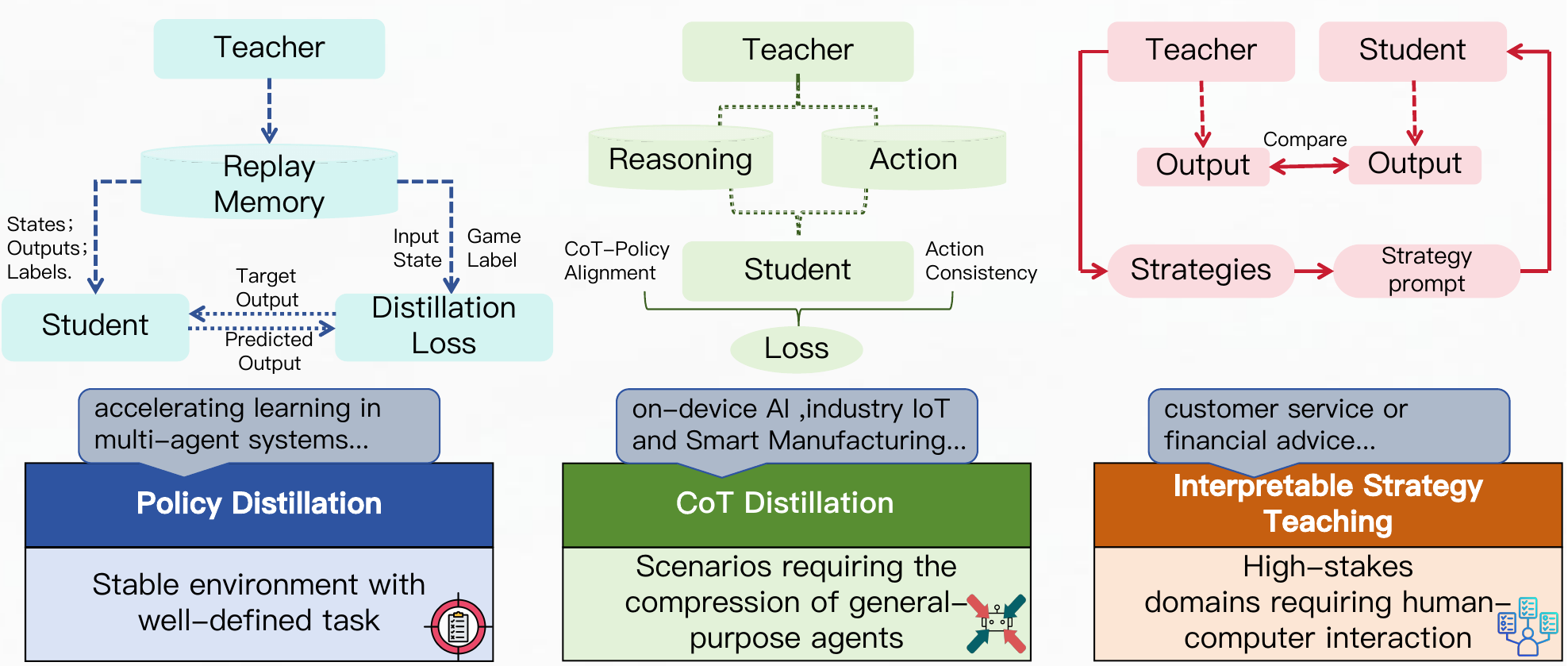}
\caption{Strategic Recommendations of Policy Distillation, Interpretable Strategic Teaching and CoT-Based KD. Policy Distillation for stable environments with well-defined tasks \cite{PKD}; Interpretable Strategy Teaching for high-stakes domains requiring human-computer interaction \cite{int}; CoT-Based KD for scenarios requiring the compression of general-purpose agents \cite{CoTKD}}
\label{Strategic}
\end{figure}

Based on the analysis above, the following strategic recommendations can be provided for different application scenarios in Fig.~\ref{Strategic}.

\subsubsection{Capability in ``Action" }

In the context of EGI, action represents the system’s capacity to translate perception and reasoning into concrete operations within its environment. It encompasses both digital and physical interactions, ranging from executing database queries or invoking local tools to coordinating with external devices or robotic actuators. KD transcends its origin as a mere model compression technique, evolving into a fundamental mechanism that endows EGI with a suite of critical, action-oriented capabilities. Specifically, KD enhances \textit{performance-optimized decision making} \cite{Mem4} by guiding lightweight models toward robust and adaptive policies—for instance, enabling mobile agents to perform low-latency beam selection in wireless communications. It enriches \textit{action reliability under data scarcity} \cite{PL1} \cite{CSI2} through pseudo-labeling and synthetic data generation, which can stabilize handover decisions when limited user trajectory data are available. It enables \textit{cross-domain and multi-task generalization}, ensuring that actions remain transferable and coordinated across heterogeneous settings such as UAV-assisted networks\cite{UAVN} and RIS-enabled environments \cite{Ac6,RISe}. Finally, it facilitates \textit{privacy-preserving, efficient coordination} among distributed agents, exemplified by federated KD frameworks for collaborative spectrum sensing or interference management. Together, these dimensions illustrate how KD serves as a unifying toolkit to strengthen the effectiveness, generalizability, and trustworthiness of actions in EGI.

\paragraph{\bf Performance-Optimized Decision Making}
Traditionally, decision-making at the edge is constrained by limited compute and stringent latency requirements, often forcing agents to trade off accuracy for efficiency. Contrary to this limitation, recent developments in KD show that lightweight students can even outperform their teachers, as demonstrated in Born-Again Networks \cite{ref7} and self-distillation frameworks \cite{ref8}. Mechanistically, KD regularizes the optimization trajectory toward flatter minima \cite{ref9,ref10}, which enhances robustness under dynamic and uncertain environments. Moreover, extensions such as bias-corrected distillation \cite{ref14,ref15} mitigate flawed teacher guidance, enabling edge agents not only to replicate but also to refine decision policies. For EGI, this transforms action from resource-limited execution into adaptive and resilient control in real-time contexts.

\paragraph{\bf Action Enrichment through Data Augmentation}
EGI agents frequently work under data-scarce or noisy conditions, where direct supervision or high-quality data is difficult to obtain. KD addresses this limitation by leveraging soft targets, pseudo-labeling, and synthetic data generation to enrich training signals, thereby enabling action policies to remain reliable even in uncertain scenarios. In data-scarce settings, Li \textit{et al.} \cite{ref16} and Wu \textit{et al.} \cite{ref17} demonstrated that KD serves as a powerful amplifier of data value, primarily through pseudo-labeling. The teacher’s soft outputs provide richer, more regularized signals than one-hot labels, making KD particularly effective in semi-supervised and unsupervised learning \cite{Kang}. Beyond generic vision or language tasks, such mechanisms are also promising in wireless applications. For instance, pseudo-labeling can support automatic modulation classification under limited labeled samples, enhance channel state information feedback with synthetic training data, or improve spectrum sensing reliability in low-SNR environments.\cite{Ac7} Mechanistically, techniques such as confidence thresholding and explicit denoising frameworks \cite{ref16,co} enhance robustness against noisy supervision. Beyond pseudo-labeling, generative models such as GANs, a class of models in which a generator and a discriminator are trained in opposition to produce realistic synthetic samples, have been employed to synthesize new data, which are subsequently labeled by the teacher to train the student, aligning with the paradigm of Data-Centric AI that emphasizes improving data quality over altering model architectures \cite{ref19}. For EGI, these advances translate into action policies that can maintain confidence and adaptability under severe data scarcity, strengthening agents’ ability to act effectively in unpredictable edge environments.

\paragraph{\bf Cross-Domain and Multi-Task Actions}
In real-world EGI systems, actions often span heterogeneous domains such as communication, sensing, and control, requiring policies that generalize across tasks and modalities. KD has emerged as a powerful enabler of such cross-domain and multi-task action integration. A central challenge lies in the heterogeneity of feature spaces, which KD mitigates through feature space alignment techniques \cite{ref20,ref21}, often enhanced by contrastive learning strategies \cite{ref22}. In multi-task learning (MTL), KD addresses negative transfer and asynchronous convergence by facilitating knowledge sharing across tasks \cite{ref23}. For example, Li \textit{et al.} \cite{ref24} proposed leveraging multiple task-specific expert teachers to guide a unified multi-task student, thereby balancing specialization and generalization. More broadly, distilled knowledge can function as a shared intermediate representation, allowing diverse models to be modularly composed into larger AI systems. In wireless networks, these advances enable models to transfer knowledge across heterogeneous tasks, such as modulation classification \cite{Cross1}, channel estimation \cite{ES3}, and interference management, thereby enhancing adaptability and scalability in dynamic communication environments.

\paragraph{\bf Privacy-Preserving and Efficient Action Coordination}
EGI relies on distributed agents collaborating across heterogeneous devices and networks, where privacy protection and communication efficiency are critical. KD provides an effective solution by enabling the exchange of distilled outputs (e.g., logits) instead of raw parameters or sensitive data, thereby reducing both privacy risks and communication overhead. Li \textit{et al.} \cite{ref26} demonstrated that output-sharing on a public proxy dataset enhances privacy \cite{ref27} while improving communication efficiency \cite{ref28}, a significant improvement over standard federated averaging (FedAvg) \cite{ref25}. However, dependence on public datasets introduces new challenges such as dataset selection and potential leakage. To address this, Lu \textit{et al.} \cite{FD} and Zhu \textit{et al.} \cite{FD2} employed generative models to synthesize proxy data, avoiding reliance on external datasets. Further advances include decentralized peer-to-peer (P2P) and co-distillation frameworks \cite{DC,P2P,P2P2}, where clients directly exchange knowledge with neighbors \cite{DC2}, often using local data as a temporary proxy for peer evaluation and weighted updates \cite{W}. For EGI, these approaches enable privacy-preserving and resource-efficient action coordination, allowing edge agents to collaborate securely and effectively in distributed environments.

\subsubsection{Capability in ``Memory" }

In the context of EGI, memory can be understood across three complementary dimensions. Model memory concerns the ability to store compressed or distilled models on resource-constrained edge devices, reflecting how efficiently knowledge is represented and retained for long-term use. Operational memory refers to the dynamic allocation of memory during inference and training, which directly impacts the agent’s capacity to execute real-time tasks under limited hardware resources. Finally, knowledge memory captures the system’s capability to preserve and recall past knowledge, for example retaining information about channel quality variations \cite{Mem6}, user mobility patterns \cite{Mem7}, or interference conditions \cite{Mem8} in wireless networks, which can be leveraged in continual learning scenarios \cite{Mem5} or through the sharing of distilled global representations in federated settings.

Recent works have explored how KD can effectively enhance memory efficiency at the edge through storage reduction, runtime optimization, and knowledge retention. Chen \textit{et al.} \cite{Mem1} introduced a computation offloading framework that integrates Deep Imitation Learning with KD to reduce task latency in heterogeneous edge–cloud settings. By distilling offline-learned optimal strategies into lightweight student models, their method enables real-time decision-making on mobile devices and offers a principled alternative to heuristic offloading approaches. Li \textit{et al.}\cite{Mem2} tackled catastrophic forgetting which appears because Deep neural networks tend to forget the learned knowledge of previous tasks when sequentially trained on a stream of tasks. Unlike rehearsal-based methods that demand high storage and risk privacy leakage, their data-free distillation embeds past knowledge into gradient-based regularization, allowing models to recall prior tasks without retaining raw data. This shift highlights KD’s role in strengthening functional memory for lifelong edge learning. Extending from knowledge preservation to distributed personalization, Pan \textit{et al.}\cite{Mem3} introduced FedCache 2.0, where dataset distillation replaces parameter exchange. Clients contribute compact synthetic datasets to a central cache, reducing communication and storage burdens while enabling adaptive personalization, though sensitivity to client heterogeneity remains a challenge.

Together, these works illustrate a layered trajectory: from optimizing runtime memory, to safeguarding knowledge memory, to redefining storage and communication memory, and finally to enhancing functional memory and robustness. KD thus emerges as a versatile paradigm for strengthening memory in diverse edge intelligence scenarios, forming a natural bridge toward the broader vision of Edge General Intelligence.

\begin{table*}[h!]
\centering
\footnotesize
\renewcommand{\arraystretch}{1.4}
\caption{A Comprehensive Comparative Analysis of Sequence Modeling Architectures:}
\label{tab:arch_comparison}
\scriptsize
\begin{tabular}{m{2cm}|m{3.5cm}|m{3.5cm}|m{3.8cm}|m{3.4cm}}
\hline
\hline
\rowcolor{lightskyblue}
\textbf{Feature} & \textbf{Mamba} & \textbf{RWKV} & \textbf{Transformer} & \textbf{RNN (LSTM/GRU)} \\ 
\hline
\textbf{Core \newline Mechanism} & Selective State Space Model & Linear Attention with Time Decay & Self-Attention & Gated Recurrence\\ 
\hline
\rowcolor{softorange}
\textbf{Training Time Complexity} & $O(B \cdot L \cdot D \cdot N)$ & $O(L \cdot d^2)$ & $O(L^2 \cdot d)$& $O(L \cdot d^2)$\\
\hline
\textbf{Training Space Complexity} & $O(B \cdot L \cdot D \cdot N)$ & $O(L \cdot d^2)$& $O(L^2 + L \cdot d)$& $O(d^2)$\\ 
\hline
\rowcolor{softorange}
\textbf{Inference Time (per token)} & $O(1)$:Constant time & $O(1)$:Constant time & $O(L)$:Linear in sequence length & $O(1)$:Constant time \\ 
\hline
\textbf{Inference Space (state)} & $O(1)$:Constant space.& $O(1)$:Constant space. & $O(L)$:Linear in sequence length& $O(1)$:Constant space. \\ 
\hline
\rowcolor{softorange}
\textbf{Key \newline Advantages}&$\bullet$ State-of-art performance\newline$\bullet$ Linear time complexity \newline$\bullet$ Extremely fast inference\newline $\bullet$ Excellent long-context scaling&$\bullet$ Excellent inference efficiency\newline$\bullet$ Parallelizable training\newline$\bullet$ No quadratic bottleneck \newline $\bullet$ Simple recurrent formulation&$\bullet$ Best-in-class performance at moderate scales \newline$\bullet$ Parallelizable training \newline $\bullet$ well-established ecosystem \newline$\bullet$ No information bottleneck&$\bullet$ Extremely low inference compute cost \newline$\bullet$ Conceptually simple\newline$\bullet$ Memory footprint during inference\\ 
\hline
\textbf{Key \newline Disadvantages} & $\bullet$ Less explored architecture \newline $\bullet$ May underperform LTI SSMs on continuous data \newline $\bullet$ Relies on custom CUDA kernels for efficiency & $\bullet$ Information bottleneck limit performance on tasks requiring high-fidelity recall \newline $\bullet$ Highly sensitive to prompt structure/information order & $\bullet$ Quadratic scaling issues in time and memory \newline $\bullet$ Slow and memory-intensive inference due to KV cache\newline$\bullet$ Prohibitively expensive for very long sequences&$\bullet$ Really difficult to parallelize training \newline$\bullet$ Vanishing gradient problem\newline$\bullet$ Poor at capturing long-range dependencies\\ 
\hline
\rowcolor{softorange}
\textbf{Applications in \newline Wireless area} & $\bullet$ General-purpose applications in wireless communications and networking \cite{MWL,MWL3}\newline $\bullet$ Lightweight, efficient real-time human activity and gesture recognition with wireless signals \cite{MWL1}   & $\bullet$ Detecting malicious node attacks in wireless sensor networks under harsh environmental conditions \cite{RWL} \newline $\bullet$ Channel estimator for Vehicle-to-Everything communications \cite{RWL2}& $\bullet$ Vision Transformer-based semantic communication for efficient and robust image transmission \cite{TWL} \newline $\bullet$ Deep learning-based hybrid beamforming in millimeter-wave MIMO-OFDM systems \cite{TWL1} &$\bullet$ Proactive resource optimization in optical backbone networks through deep learning-based traffic prediction \cite{RNWL}\newline$\bullet$ Low-complexity CSI prediction in massive MIMO (mMIMO) systems \cite{RNWL1}\\ 
\hline
\multicolumn{5}{c}{L: sequence length\qquad D(d): model dimension\qquad      B: batch size\qquad      N: state dimension}\\
\hline
\hline
\end{tabular}
\end{table*}
\subsection{Beyond Transformers: Architectures Natively Designed for the Edge}

The inherent quadratic complexity of the Transformer architecture remains a fundamental bottleneck for edge deployment. This has spurred the development of novel architectures engineered for computational efficiency. However, these emerging models, such as Mamba\cite{ref49} and RWKV \cite{ref60}, often lack the extensive pre-training on vast datasets that underpins the powerful capabilities of their Transformer-based predecessors. KD, particularly in the form of cross-architecture distillation, emerges as a critical and capital-efficient strategy to bridge this gap. By transferring the rich, generalized knowledge from a large, pre-trained Transformer "teacher" to a more efficient "student" model like Mamba or RWKV, it is possible to imbue these lightweight architectures with advanced capabilities without incurring the prohibitive costs of training them from scratch. This symbiotic relationship between efficient novel architectures and sophisticated distillation techniques is paving the way for the next generation of deployable, high-performance models for Edge General Intelligence.

\subsubsection{Mamba: Architecture with Linear-Time Sequence}

The Mamba architecture can be regarded as a significant evolution of the Structured State-Space Model (S4) \cite{ref49,ref50,ref51}, inheriting its mathematical formulation while introducing novel mechanisms to enhance flexibility and efficiency. Mamba also incorporates an efficient algorithm with parallel scan and kernel fusion, streamlining its architecture by unifying the separate attention and MLP layers of a Transformer into a homogeneous Mamba block \cite{ref49}. Mamba maintains the latent state formulation, where the hidden state evolves over time based on the input and parameterized dynamics.

Recent research has established the Mamba architecture as a highly efficient and versatile tool for wireless systems. For network optimization, \textit{Mehrabian and Wong} \cite{MWL3} proposed the A-Gamba model for cellular traffic prediction, achieving a 7$\times$ faster training time and a 681$\times$ reduction in multiply-accumulate operations (MACs) over strong baselines . In core communication functions, \textit{Zhang et al.} \cite{MWL} demonstrated that Mamba can enhance semantic decoding, while also enabling scalable resource allocation such as energy-efficient beamforming with near-constant inference time as user numbers grow. Moreover, in wireless human sensing, \textit{Liu et al.} \cite{MWL4} developed TF-Mamba that reduces MACs by up to 98.9\% while achieving high accuracy. Similarly, \textit{Huang et al.} \cite{MWL1} created SenseMamba, which achieves state-of-the-art accuracy with only 0.021M parameters. Extending this to multimodal perception, \textit{Yang et al.} \cite{MWL5} showed that a Mamba-based framework could improve performance when sensor data is missing by fine-tuning less than 0.3\% of its parameters. Collectively, these attributes underscore Mamba's potential as a key enabling technology for decentralized, edge-native intelligence in next-generation wireless networks and even the possible way towards mobile agentic AI.

\subsubsection{RWKV: RNN-Transformer Hybrid Architectures}

The core purpose of RWKV \cite{ref60} is to integrate the high performance of Transformers with the computational efficiency of recurrent neural networks (RNNs). Its core innovation lies in a linear-complexity attention mechanism: the hidden state \(h(t)\) depends only on \(h(t-1)\) and the current input \(x(t)\), enabling step-by-step computation with constant memory and linear time complexity.

RWKV is composed of four fundamental components: (i) R(Receptance): a gating mechanism regulating how much past information influences the current state. (ii) W(Weight):a channel-specific, learnable time decay vector. (iii) K(Key): analogous to attention keys, computed via linear interpolation of current and past tokens. (iiii) V(Value): analogous to attention values, also derived through linear interpolation of current and past tokens. Together, these define the WKV operator, which unifies transformer-style parallel training with RNN-style efficient inference.

Recent studies have leveraged the RWKV architecture to create highly efficient solutions. For instance, \textit{Fu et al.} \cite{RWL2} developed an RWKV-based channel estimator for Vehicle-to-Everything communications that reduces computational complexity by 24.9\% compared to LSTM-based methods. In the domain of network security, \textit{Xie et al.} \cite{RWL} proposed an RWKV-based scheme to detect selective forwarding attacks in wireless sensor networks, achieving an average false detection rate of just 0.09\% in harsh mobile environments. Critically, the architecture's suitability for the wireless edge has been bolstered by the work of \textit{Choe et al.} \cite{RWL3}, who introduced a suite of compression techniques that reduce the RWKV model's memory footprint by 3.4$\times$ to 5$\times$ with only negligible accuracy degradation, making it feasible for resource-constrained devices.

\subsubsection{Cross-Architecture Distillation}:

New architectures such as Mamba and RWKV promise superior efficiency but lack the trillions of tokens of pre-training data and massive computational investment that have made Transformers so capable \cite{T2S}. Cross-architecture distillation breaks this cycle by providing a capital-efficient pathway \cite{LL}. 

Unlike same-architecture distillation, heterogeneous architectures exhibit significant feature divergence \cite{CKD1}.This mismatch is particularly acute when distilling non-causal models like Transformers into causal models like Mamba. Consequently, naive feature-based distillation methods often fail, necessitating more sophisticated alignment techniques \cite{CKD2}. To overcome these challenges, Yao \textit{et al.} \cite{T2M} constructed a hybrid model composed of 86\% of Mamba blocks and 14\% of self-attention blocks, leveraging the computational efficiency of Mamba blocks while strategically incorporating self-attention blocks to handle global interactions. Another direction in cross-architecture distillation involves the use of specially designed ``projectors" to align feature spaces. Liu \textit{et al.} \cite{CKD} and Hao \textit{et al.} \cite{CKD2} introduced methods that utilize these projectors to map the student's features into the teacher's feature or attention space. 

As discussed previously, the high efficiency and less model parameters achieved through cross-architecture distillation make it a highly advantageous technique for edge deployment. To enable the deployment of an accurate retinal disease classifier on an edge device, \textit{Yilmaz and Aiyengar} \cite{CKDWL} used a cross-architecture distillation framework to transfer diagnostic knowledge from a large Vision Transformer teacher model to a lightweight CNN student model. Their framework, featuring specialized projectors, dramatically improved the student's performance with 97.4\% reduction in the model parameters (from 85.8M to 2.2M), which resulted in a compact 8.79 MB model that retained 93\% of the teacher's diagnostic accuracy on the edge device. The primary strength of this approach is the successful creation of an efficient, high-accuracy model for resource-constrained settings. However, its limitations include being trained on a relatively small dataset and for a limited number of epochs due to computational constraints \cite{CKDWL}.

\subsection{Utilize KD to obtain compact and deployable edge models}
\subsubsection{Language Models as a Foundation: From BERT to TinyBERT}
Language models (LMs) are central to the core functions of modern intelligent agents, operationalizing intelligence as the capability to interact with environments, utilize tools, and achieve goals. However, the prevailing architecture, which relies on Large Language Model (LLM) API endpoints hosted on centralized cloud infrastructure, is fundamentally incompatible with the resource-constrained nature of the edge due to computational cost, memory footprint, and energy consumption. 

Consequently, a new vision is emerging where general intelligence is realized not as a singular cloud entity but as the aggregate capability of numerous specialized SLMs-powered agents operating autonomously in local contexts. The true potential of SLMs for edge intelligence is unlocked by knowledge distillation. In this section, we use BERT as a case study to illustrate this point.

The deployment challenges posed by large-scale models like Bidirectional Encoder Representations from Transformers (BERT), whose exceptional performance is linked to its immense size, are systematically addressed by knowledge distillation. 

Sanh \textit{et al.} \cite{DBERT} create a miniaturized general-purpose version of BERT, which serves as a seminal example of successfully applying knowledge distillation. Its core contribution lies in demonstrating that distillation can be performed during the computationally intensive pre-training phase. DistilBERT's compression effects are remarkably significant, successfully striking an excellent balance between model efficiency and performance.Compared to BERT-Base, DistilBERT reduces the parameter count by 40\% and achieves a 60\% increase in inference speed.Moreover, DistilBERT retains 97\% of BERT-Base's language understanding capabilities. On the GLUE benchmark, which includes 9 different tasks, its average score is very close to that of BERT-Base.

\begin{figure*}[htbp]
    \centering
    \includegraphics[width=0.9\textwidth]{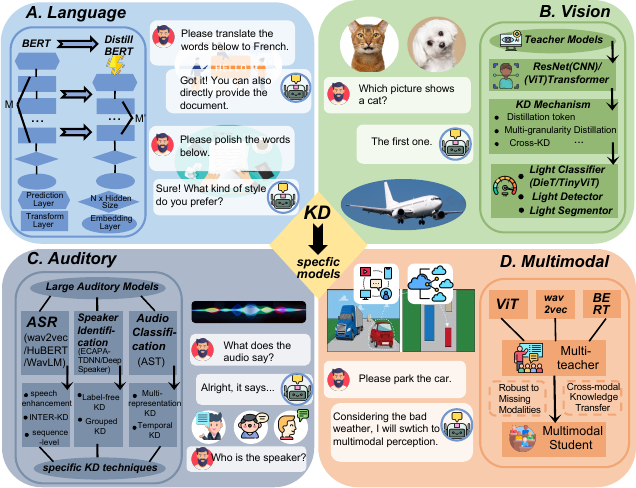} 
    \caption{An overview of KD techniques across different modalities. (A) Language: distillation from large language models such as BERT into compact models (e.g., DistilBERT). (B) Vision: KD mechanisms including distillation tokens, multi-granularity distillation, and cross-KD applied to light classifiers, detectors, and segmentors. (C) Auditory: specific KD strategies such as sequence-level KD, label-free KD, grouped KD, multi-representation KD, and temporal KD for tasks like ASR, speaker identification, and audio classification. (D) Multimodal: integration of KD across modalities with multi-teacher settings, robust handling of missing modalities, and cross-modal knowledge transfer for multimodal student models.}
\end{figure*}
If DistilBERT represents the pioneering application of KD to BERT,then Jiao \textit{et al.} \cite{TBERT} embodies a more profound and refined distillation strategy. TinyBERT pioneers a ``Transformer distillation" method that delves deep into the internal mechanisms of the Transformer, aiming for a fine-grained imitation of the teacher model's intermediate representations at every layer. 

TinyBERT has achieved astonishing results in model compression with its advanced two-stage, multi-level distillation framework.
Taking the 4-layer TinyBERT4 as an example, its parameter count is merely 14.5 million. Compared to BERT-Base, its model size is reduced by 7.5 times, while its inference speed is boosted by 9.4 times. While achieving such extreme compression, TinyBERT4 retains 96.8\% of BERT-Base's performance on the GLUE benchmark, with minimal performance loss. Even more impressively, a 6-layer TinyBERT6 performs nearly on par with the full BERT-Base teacher model on GLUE. The emergence of TinyBERT provides a powerful and reliable option for deploying high-performance NLP models designed to process and understand human language on edge devices with extremely limited resources.

\subsubsection{Distilled Models for Wireless Tasks}

In wireless communication scenarios, the development of compact distilled models has not yet reached the same level of standardization as in NLP, where models such as DistilBERT and TinyBERT serve as well-established benchmarks. Nevertheless, in several wireless-specific tasks—such as CSI feedback and reconstruction, there exist a variety of efficient and task-tailored knowledge distillation frameworks specifically designed to address the unique characteristics of wireless systems.

\begin{itemize}
\item \textbf{CRNet-SE \cite{CSI2}}
    
In recent years, various KD-based models have been proposed to enhance CSI feedback and edge inference efficiency in wireless communication systems. Among them, CRNet-SE focuses on the CSI feedback problem in large-scale MIMO systems. Its main objective is to compress high-dimensional CSI on user equipment (UE) with limited computational resources. The teacher model, CRNet, employs a multi-resolution encoder to extract rich channel features, while the student model simplifies the encoder into only one convolutional and one fully connected layer for lightweight deployment. Despite this reduction, the KD-trained student achieves a significantly lower NMSE than a non-distilled counterpart, illustrating the efficacy of KD in preserving reconstruction accuracy under severe model compression.

\item \textbf{CSI-ALM-Light \cite{CSI-ALM-Light}}

While CRNet-SE emphasizes compression for feedback efficiency, CSI-ALM-Light extends the KD paradigm toward temporal prediction in high-mobility communication scenarios. Instead of reconstructing current CSI, CSI-ALM-Light predicts future downlink CSI using historical uplink information. Its teacher, CSI-ALM, builds on a pretrained GPT-2 backbone and integrates a Modality Alignment Mechanism to align numerical CSI features with language embeddings. The distilled student replaces these complex components with a lightweight Transformer and learnable soft prompts, while employing a self-attention relation–based distillation strategy that aligns multi-level attention relations (Query–Key, Query–Query, Key–Key, and Value–Value). Remarkably, CSI-ALM-Light achieves near-teacher performance using only 10\% of the data and less than 1/200 of the parameters, highlighting KD’s potential to bridge performance gaps across drastically different model scales.

\item \textbf{KD-Based AIoT Framework \cite{Gesture}}

Complementary to these communication-oriented designs, the KD-Based AIoT Framework applies KD to Wi-Fi CSI–based gesture recognition for intelligent edge devices. Unlike CRNet-SE and CSI-ALM-Light, which focus on CSI reconstruction or prediction, this framework leverages KD to enable efficient human activity recognition (e.g., waving, walking, falling) under strict memory and computation constraints. The student network, containing merely 28K parameters, integrates multiple parallel convolutional branches and a bottleneck fusion layer, and through a joint response-plus-feature distillation scheme, it achieves up to 99.5\% accuracy on SignFi datasets—outperforming both response-only and hard-label training.
\end{itemize}

Overall, these studies collectively demonstrate that KD serves as a versatile mechanism across diverse wireless and AIoT tasks—from channel compression (CRNet-SE), to temporal prediction (CSI-ALM-Light), and edge activity recognition (KD-Based AIoT)—each highlighting a distinct yet complementary facet of efficient model design under hardware constraints. 

\subsection{Lesson learned}

A lesson from this analysis is that KD is not a uniform methodology but a collection of adaptive strategies whose effectiveness varies by the target agentic capability. In NLP, KD primarily preserves architectural fidelity and semantic representation, while in wireless communication tasks, domain constraints such as limited data and noisy, resource-restricted edge environments shift the focus toward task-oriented adaptation, as demonstrated by models and the KD-Based AIoT Framework \cite{Gesture}. However, existing wireless KD approaches remain mostly task-specific prototypes and lack generalized distillation protocols comparable to those in NLP \cite{Mem4}. Moreover, KD efficiency depends heavily on the teacher’s domain adaptability and the quality of intermediate representations \cite{Lesson2,Lesson3}, which are harder to define for non-textual modalities. Future directions include establishing standardized KD benchmarks for edge wireless intelligence, developing modality-agnostic distillation mechanisms transferable across heterogeneous inputs \cite{Lesson4}, and enabling collaborative KD across multiple edge nodes to approximate cloud-level intelligence \cite{Lesson1}.

\section{Application}\label{Four}
This section explores the specialized architectures and technologies for specific domains such as Unmanned Aerial Vehicles (UAVs), Autonomous Vehicles(AV), Robotics and Other IoT Devices. 

\subsection{UAV}

KD is a critical technique for deploying complex models on resource-constrained UAVs, with researchers developing specialized frameworks to enhance on-device efficiency and performance across different operational domains. These approaches can be broadly categorized by their primary objective, whether optimizing high-level mission logistics, bolstering on-board cybersecurity, or refining the fundamental distillation process for perception tasks.

To begin with,in the domain of mission optimization, \textit{Sun et al.} \cite{UAVKD} created a cooperative co-evolutionary framework to address the challenge of running complex task-planning algorithms on energy-limited UAVs. In their approach, multiple student subnetworks collaboratively learn and exchange information to compress a large Deep Neural Network (DNN) to just 1.3\% of its original parameters. The use of KD here drastically reduces task completion delay by up to 95\% in large-scale scenarios. This work illustrates that efficiency is only one side of the equation, security remains another vital concern for autonomous UAVs operating in dynamic and adversarial environments.

The efficiency gains from KD are foundational, not merely incremental, for the evolution of UAVs. By enabling complex cognitive capabilities such as on-device reasoning and planning, KD provides the crucial bridge to transform UAVs from automated tools into autonomous, goal-driven Agentic UAVs \cite{UAV1}. This latest evolution, the Agentic UAV, combines cognitive AI models, multimodal sensing, and edge computing to support real-time perception and reasoning. To address cloud latency and limited onboard resources, Agentic UAVs employ compact edge AI platforms for tasks such as semantic segmentation and path reconfiguration \cite{UAV1}. The capabilities of these agents are further extended by Vision-Language Models (VLMs), which enable the execution of high-level user commands through semantic grounding and zero-shot generalization. Thus, KD serves as the enabling foundation upon which higher-order autonomy and cognitive generalization are being built in modern UAV systems.

However, the story of UAV intelligence does not end at the single-agent level. Given the persistent limitations on battery and processing power\cite{UAV-y}, UAVs often offload compute-intensive tasks, such as 3D mapping and large-scale inference, to edge servers \cite{UAV2}. This shift from individual to distributed intelligence marks the transition toward swarm-based and EGI-powered UAV ecosystems, where multiple agents collaboratively sense, reason, and act. In summary, while edge computing techniques like KD greatly enhance UAV autonomy by overcoming inherent physical limitations, the evolution toward distributed cognitive systems like EGI-powered swarms introduces new challenges. The opacity of deep learning creates an ``explainability-accountability crisis," complicating error analysis, while physical safety remains a critical, underexplored area for ensuring public acceptance and safe deployment \cite{UAV2}.

\begin{figure*}
    \centering
    \includegraphics[width=0.6\textwidth]{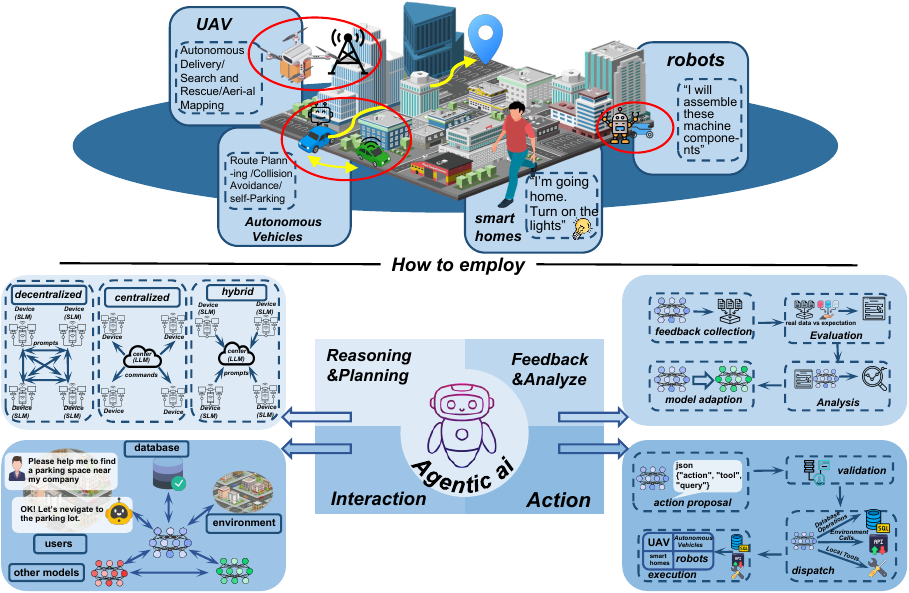}
    \caption{Illustration of KD in enabling EGI within mobile agentic AI ecosystems. Use cases span UAVs (e.g., autonomous delivery, search and rescue, aerial mapping), autonomous vehicles (e.g., route planning, collision avoidance, self-parking), robots (e.g., task assembly, household assistance), and smart homes. The figure highlights how agentic AI leverages both LLMs in the cloud and smaller student models (SLMs) at the edge to achieve reasoning, planning, feedback, and interaction. KD facilitates the transfer of memory and learned knowledge, enabling mobile agentic AI to perform efficient, real-time, and context-aware decision-making at the edge.}
\end{figure*}
\subsection{Autonomous Vehicles}
Achieving Level-5 autonomous driving, where a vehicle can operate in all environments without any human intervention, requires extremely robust, low-latency, and high-reliability AI capabilities. Such systems must continuously process large volumes of high-dimensional, multi-modal sensor data (e.g., camera, LiDAR, radar, GPS, IMU) and make real-time decisions for perception, planning, and control \cite{ref134}.
 
Current approaches to supporting autonomous vehicles can be broadly categorized into five types: vehicle-based, vehicle-to-vehicle collaborative computing, cloud-assisted, edge-assisted, and cloud-edge-assisted approaches \cite{ref135}.To realize EGI for autonomous vehicles (AVs), general methods like KD are complemented by specialized techniques such as Bird’s-Eye-View (BEV) perception and Collaborative Edge Intelligence (CEI).

BEV perception transforms and fuses data from multiple cameras, LiDAR, or millimeter-wave radar sensors to generate a unified, top-down spatial representation. This approach mitigates challenges like perspective distortion and occlusion inherent in single sensors, providing downstream path planning systems with more stable and information-rich inputs. 

CEI is a distributed computing paradigm that integrates geo-distributed edge resources into a federated pool. It enables both vertical collaboration (cloud-edge-end) and horizontal collaboration (edge-to-edge), creating a decentralized intelligent network that can dynamically orchestrate resources to meet the stringent demands of autonomous driving \cite{ref135}.
   
\subsection{Robotics}
The strict real-time and security requirements of robotic Operational Technologies (OT) are often poorly met by centralized cloud computing, making Edge Computing a critical complementary approach that provides computation closer to the system \cite{ref139}. Several frameworks facilitate this, including FogROS2 \cite{ref140} for task offloading, the EI-for-AMR SDK \cite{ref142} for containerized deployment and hardware-accelerated inference, and the Edge-Driving Robotics Platform (EDRP) \cite{ref143} for 5G-enabled collaborative mapping.

General-purpose robotics often involves perceiving and manipulating complex objects. Several studies have sought to address such challenges. For example, Chen \textit{et al.} \cite{ref144} adopt an innovative teacher-student learning paradigm, in which a 'privileged agent' equipped with complete information about the state of the cloth serves as the teacher, guiding the robot through complex cloth manipulation tasks using vision-based reinforcement learning. Similarly, for high-precision industrial tasks, Zhao \textit{et al.} \cite{ref145} proposes a novel semi-supervised knowledge distillation framework aimed at enhancing the robustness and accuracy of vision-based guidance in intelligent manufacturing settings.

Policies optimized in simulation often fail when deployed on physical robots due to discrepancies between simulated and real-world dynamics. A common solution involves training a teacher policy in simulation with access to privileged information (e.g., precise object velocities, friction coefficients), while the student policy learns to replicate its behavior using only the sensor data available to a physical robot (e.g., visual input, joint encoder readings) \cite{ref146}.

To manage diverse skills or collaborative multi-robot scenarios, specialized expert teacher policies can be trained for individual tasks. KD is then used to consolidate their capabilities into a unified student network. This process distills the collective behaviors of all teachers, allowing the student to learn cross-task commonalities and form a compact, generalizable decision-making model \cite{ref147,ref133}.

 \subsection{Other IoT Devices}
The Internet of Things(IoT) features a massive, growing network of interconnected devices that generate immense volumes of heterogeneous, real-time data \cite{IoT1}.

The Internet of Agents (IoA) provides the architectural blueprint to realize EGI, offering a scalable, agent-centric infrastructure. Its hierarchical, layered design is engineered to manage the complexity of a global agent network \cite{IoT2}. The foundational infrastructure layer supplies essential resources, including AI models, diverse computing environments, multimodal data, and high-reliability communication networks.

Effective multi-agent communication in IoT environments requires balancing efficiency, reliability, and semantic richness. While lightweight protocols like MQTT \cite{IoT3} and CoAPare efficient for simple sensor data, they lack the semantic depth for complex agent interactions \cite{IoT4}. To address this, emerging agent-centric standards like Google’s Agent-to-Agent (A2A) protocol and Anthropic’s Model Context Protocol (MCP) enable goal-oriented collaboration. A2A uses ``Agent Cards” for dynamic discovery and task coordination, while MCP unifies access to external tools and real-time data to enhance contextual awareness \cite{IoT5}.

Agentic AI and the IoA are transitioning from theory to practice, with applications emerging across several domains:
\begin{itemize}
\item \textbf{Smart Homes}: Systems like Sasha \cite{IoT6} and SAGE \cite{IoT7} demonstrate advanced goal reasoning and personalization, while decentralized device-as-agent architectures enable proactive, privacy-preserving collaboration at the edge \cite{IoT2}.
\item \textbf{Urban and Industrial IoT}: 
Frameworks such as CityGPT \cite{IoT8}and CASIT \cite{IoT10} showcase large-scale multi-agent coordination. CityGPT augments human decision-making with spatiotemporal data analysis,and CASIT autonomously manages remote environmental monitoring by using hierarchical agent structures to optimize bandwidth.\cite{IoT14}
\item \textbf{Smart Transportation}: Connected and Autonomous Vehicles (CAVs) \cite{IoT11} operate as cooperative agents via V2V and V2I communication \cite{IoT13}for applications like cooperative merging and platooning. Internally, multi-agent hierarchies manage planning, tactical decisions, and energy optimization.
\end{itemize}

\subsection{Lesson learned}

Across domains such as UAVs \cite{UAV1}, autonomous vehicles \cite{ref134}, robotics \cite{ref142}, and IoT devices \cite{IoT1}, KD consistently enables EGI by compressing large teacher models into lightweight, deployable student models for resource-constrained environments \cite{KDS}. These applications highlight KD’s role in bridging high-capacity models and edge agents, supporting cross-modal knowledge transfer and real-time, context-aware decision-making \cite{Les2}. However, the opacity of distilled models raises explainability and accountability concerns \cite{Les3}, and KD alone does not address physical safety in critical systems \cite{KDS}. Promising directions include safety-aware distillation and extending KD from policy transfer to semantic understanding \cite{Les4}, enabling more transparent, robust, and efficient multi-agent coordination.

\section{Challenges and Future directions}\label{Five}
\subsection{Challenges}

While these lessons highlight KD’s potential in enabling edge general intelligence, they also reveal a deeper truth: significant technical and ethical challenges remain before EGI can be deployed safely and reliably at scale{\footnote{The authors would like to thank the Qwen, an advanced large language model, for its valuable academic polish service.}}.
\subsubsection{Confronting the Technical Challenges of Edge General Intelligence}
\begin{itemize}
    \item\textbf{Toward New Benchmarks for On-Device Agentic Performance:} Most existing AI benchmarks are cloud-centric and focus only on static task accuracy, making them unsuitable for evaluating mobile agentic AI at the edge. Current edge benchmarks suffer from limited hardware and software coverage, unrealistic testing environments, lack of multi-tenancy support, and weak end-to-end evaluation. Moreover, they typically measure accuracy but ignore latency, energy efficiency, and robustness. Traditional benchmarks like ImageNet \cite{ref104} and SQuAD \cite{ref105} rely on pre-collected datasets, while on-device agents must perform within dynamic, interactive feedback loops.
    
    \item \textbf{Mitigating Hallucinations and Errors in Compressed Agents:} Large models often hallucinate, producing confident but incorrect outputs because training with hard labels enforces deterministic supervision and overconfidence. KD alleviates this by using softened teacher distributions, but still suffers from exposure bias: teacher-generated examples often exhibit high perplexity for the student, making adaptation difficult and reducing the effectiveness of distillation \cite{ref109}.
  
  \item\textbf{Vulnerabilities in Distillation and Federated Systems:} In federated distillation (FD), the server cannot observe how clients generate logits, making logits-poisoning attacks (LPA) particularly effective. Although FDLA \cite{ref112} introduces a tailored LPA, its impact across heterogeneous clients and different distillation stages remains insufficiently studied. More sophisticated variants, such as Peak-Controlled FDLA (PCFDLA), manipulate peak logits to evade detection. Beyond these targeted attacks, Byzantine clients may degrade performance through parameter tampering or high-dimensional perturbations. Other attacks, including Local Model Averaging (LMA) and Collusion-based Poisoning (CPA), further obscure adversarial behavior. Privacy is also not guaranteed—logits can leak sensitive user information through membership inference attacks (MIAs) \cite{ref115,FL-y}.
\end{itemize}

\subsubsection{Navigating the Ethical Labyrinth of Edge General Intelligence}

The technical capabilities and limitations of EGI directly give rise to a series of profound ethical challenges. This section will systematically analyze these challenges and reveal their inextricable link to the aforementioned technical difficulties.

\begin{itemize}
    \item \textbf{The Human Element:}
   Encoding complex and sometimes contradictory human values into EGI systems is an ethical challenge. Simulated empathy without real understanding risks manipulation, while rigid ethical rules fail in situations requiring nuanced judgment (e.g., the trolley problem \cite{AGI}). The goal is not to hard-code emotions or ethics, but to enable a functional analogue of human wisdom.
    \item\textbf{Algorithmic Integrity and Trust:}
   Bias, privacy risks, and errors are inherent to data-driven AI systems \cite{ref2}. Ambiguity, adversarial inputs, or overfitting may cause harmful mistakes in high-stakes settings. Meanwhile, EGI must balance privacy with the need for authentic data, and biased training sets can lead to unfair or discriminatory decisions, especially in sensitive domains such as law enforcement, defense, and healthcare.
  \item\textbf{Societal and Economic Impacts:}
  The rise of EGI may trigger large-scale socioeconomic disruption. Its development requires massive investment, while its automation capabilities threaten broad employment sectors. Traditional responses such as retraining may be insufficient—not due to a skills gap alone, but the potential obsolescence of human labor. If EGI surpasses humans in most cognitive tasks, the role of human work in economic production becomes uncertain.
\end{itemize}
\subsection{Future Directions}
\begin{itemize}
    \item Safety-aware KD frameworks that explicitly optimize for model robustness and predictable behavior in real-world deployments.
    \item Modality-agnostic distillation mechanisms capable of transferring both structural and semantic priors across heterogeneous input forms \cite{Lesson4}.
    \item Collaborative multi-agent KD, where edge devices distill and exchange knowledge to collectively approximate cloud-level intelligence \cite{Lesson1}.
    \item Transitioning from policy transfer to semantic understanding transfer \cite{Les4}, enabling transparent decision-making and reducing the explainability gap.
   \item
    Instead of replacing human judgment, embed human oversight into decision-critical loops, ensuring accountability, interpretability, and alignment with societal norms.
\end{itemize}

\section{Conclusion}\label{Seven}
The journey towards Edge General Intelligence is not merely a technical problem of model compression but a comprehensive scientific and engineering endeavor. By adopting the principled, synergistic approach detailed in this survey, which combined agent-aware capability transfer, edge-native architectures, and efficient adaptation methods, the research community can successfully bridge the deployment chasm. The challenges that lie ahead are formidable, spanning the technical domains of benchmarking and robustness to the deep socio-ethical questions of transparency, trust, and societal impact. Yet, these challenges define a clear and compelling research agenda for the coming years. This survey has charted a viable path forward, laying the groundwork for a future where intelligent, autonomous agents can operate securely, efficiently, and beneficially at the very edge of our increasingly connected digital world.

\bibliographystyle{IEEEtran}
\bibliography{mylib}

\end{document}